\documentclass[pra,twocolumn,superscriptaddress,showpacs,nofootinbibfloatfix,amsmath,amsfonts,amssymb]{revtex4-1}%
\usepackage{amsmath,amsfonts,amssymb,color}
\usepackage{amsthm}
\usepackage{leftidx}
\usepackage{graphicx}
\usepackage{xcolor}
\usepackage{dcolumn}
\usepackage{bm}
\usepackage{epstopdf}
\usepackage{epsfig}
\usepackage{environ}
\usepackage{pdfcomment}

\usepackage{float}
\usepackage[T1]{fontenc}
\usepackage[latin9]{inputenc}
\usepackage{setspace}
\usepackage{esint}

\begin{document}
	
\preprint{APS/123-QED}

\title{\textbf{Floquet topological phases with fourfold degenerate edge modes in
a driven spin-$1/2$ Creutz ladder}}

\author{Longwen Zhou}
\email{zhoulw13@u.nus.edu}
\affiliation{%
	Department of Physics, College of Information Science and Engineering, Ocean University of China, Qingdao 266100, China
}
\affiliation{Institute of Theoretical Physics, Chinese Academy of Sciences, Beijing 100190, China}

\author{Qianqian Du}
\affiliation{%
	Department of Physics, College of Information Science and Engineering, Ocean University of China, Qingdao 266100, China
}

\date{\today}

\begin{abstract}
Floquet engineering has the advantage of generating new phases with
large topological invariants and many edge states by simple driving
protocols. In this work, we propose an approach to obtain Floquet
topological phases with fourfold degenerate edge states and even-integer topological
invariants in a spin-$1/2$ Creutz ladder model, which is realizable
in current experiments. Putting the ladder under periodic quenches,
we found rich Floquet topological phases in the system, which belong
to the symmetry class CII. Each of these phases is characterized by
a pair of even integer topological invariants $(w_{0},w_{\pi})\in2\mathbb{Z}\times2\mathbb{Z}$,
which can take arbitrarily large values with the increase of driving
parameters. Under the open boundary condition, we further obtain multiple quartets
of topological edge states with quasienergies zero and $\pi$ in the
system. Their numbers are determined by
the bulk topological invariants $(w_{0},w_{\pi})$ due to the bulk-edge
correspondence. Finally, we propose a way to dynamically probe the
Floquet topological phases in our system by measuring a generalized
mean chiral displacement. Our findings thus enrich the family of Floquet
topological matter, and put forward the detection of their topological
properties.
\end{abstract}

%Use showkeys class option if keyword
\maketitle

\section{Introduction\label{sec:Intro}}

Floquet topological states of matter have attracted great attention
in the past decade~(see Refs.~\cite{FlqRev1,FlqRev2,FlqRev3,FlqRev4,FlqRev5} for reviews). These intrinsically nonequilibrium phases could
appear in systems subject to time-periodic driving fields~\cite{FlqRev6,FlqRev7,FlqRev8}. Theoretically,
various types of topological phases that are unique to periodically
driven systems have been discovered~\cite{FlqEg1,FlqEg2,FlqEg3,FlqEg4,FlqEg5,FlqEg6,FlqEg7,FlqEg8,FlqEg9,FlqEg10,FlqEg11,FlqEg12,FlqEg13,FlqEg14,FlqEg15,FlqEg16}, and symmetry classification schemes
for Floquet topological matter have also been proposed~\cite{SymCls1,SymCls2,SymCls3,SymCls4,SymCls5}. Experimentally, Floquet topological phases have been observed in solid state materials~\cite{SolidFTP,SolidFTP2}, cold
atoms in optical lattices~\cite{ColdAtomFTP1,ColdAtomFTP2,ColdAtomFTP3}, photonic~\cite{PhotonFTP1,PhotonFTP2,PhotonFTP3} and phononic~\cite{PhononFTP1,PhononFTP2,PhononFTP3} systems. On application side, Floquet states
have the potential of realizing setups with many topological transport
channels~\cite{TransFTP1,TransFTP2,TransFTP3} and creating new schemes of topological quantum computations~\cite{FloCompt1,FloCompt2,FloCompt3}.
In recent years, the study of Floquet topological matter has also
been extended to higher-order topological models~\cite{HOFTP1,HOFTP2,HOFTP3,HOFTP4,HOFTP5,HOFTP6} and non-Hermitian systems~\cite{NHFTP1,NHFTP2,NHFTP3,NHFTP4,NHFTP5,NHFTP6,NHFTP7,NHFTP8}.

In the engineering of Floquet topological matter, one of the essential
idea is that the periodic driving fields could induce long-range hopping
and interactions on top of the non-driven system, leading to new phases
with large topological invariants and many topologically protected edge
states~\cite{FlqEg3,FlqEg4}. This idea has been successfully applied to obtain a series of Floquet topological insulating and superconducting phases in one- and two-dimensional systems,
both Hermitian~\cite{TransFTP1,TransFTP2,TransFTP3,HOFTP1,LargeTI1,LargeTI2,LargeTI3,LargeTI4,LargeTI5,LargeTI6} and non-Hermitian~\cite{NHFTP1,NHFTP2,NHFTP3,NHFTP4}, yielding large quantized Floquet-Thouless pumps~\cite{LargeTI1,LargeTI4} and edge-state transport coefficients~\cite{TransFTP1,TransFTP2,TransFTP3}. The Floquet Majorana
edge modes obtained following this idea are further applied in a spatiotemporal
proposal of topological quantum computing~\cite{FloCompt1,FloCompt2,FloCompt3}. 

According to the periodic table of topological insulators and superconductors~\cite{Tenfold1,Tenfold2},
systems belonging to the symmetry classes AIII, BDI and CII in one-dimension are characterized by integer topological invariants, and therefore
could be engineered to obtain phases with large topological numbers
and many edge modes by the Floquet method. Indeed, the studies in Refs.~\cite{FlqEg4,NHFTP1,NHFTP2,NHFTP3,NHFTP4,LargeTI2,LargeTI3} are all based on model systems in either the symmetry class
AIII or BDI. However, systems belonging to the symmetry class CII in
one-dimension, which are characterized by even integer topological invariants
($2\mathbb{Z}$) and fourfold degenerate edge modes are seldomly been
considered in the Floquet engineering. One possible reason for such
a bias is that a one-dimensional~(1D) model in the CII class has at least four Floquet quasienergy
bands, and therefore its theoretical treatment is more complicated
than those two-band candidates in AIII and BDI classes. Nevertheless,
when Floquet driving fields are applied, the $2\mathbb{Z}$-characterization
of CII-class models imply that they could support topological phases
with even more edge modes under the open boundary condition (OBC),
and the properties of these new Floquet phases certainly deserve a
detailed study.

Motivated by the above considerations, in this work we propose a 1D
Floquet system in the symmetry class CII, whose topological phases
are characterized by a pair of even integer topological invariants
($2\mathbb{Z}\times2\mathbb{Z}$), and also featured by quartets of
topological edge modes with fourfold degeneracy at both quasienergies
zero and $\pi$.
Our construction is based on the Creutz ladder (CL) model, which is originally proposed as a toy model to study chiral fermions in lattice gauge theory~\cite{CL1,CL2,CL3}. It is a quasi-1D lattice formed by two coupled chains, populated with spinless fermions and subjected to a perpendicular magnetic field. In early studies, it was already identified that for certain values of the magnetic flux, the model belongs to the symmetry class BDI, and there are chiral symmetry protected zero-energy edge modes at the boundaries of the ladder~\cite{CL2}. The CL therefore realizes one of the earliest examples of a 1D topological insulator. In later investigations, various modified versions of the CL are proposed~\cite{CL4,CL5,CL6,CL7,CL8,CL9,CL10,CL11,CL12,CL13,CL132,CL133,CL21,CL22}, and also realized experimentally in atom-optical setups like cold atoms~\cite{CL16,CL17} and photonic lattices~\cite{CL14,CL15}. Intriguing topological features of the CL are reflected in the formation of Aharonov-Bohm cages~\cite{CL15,CL18,CL19}, localization dynamics~\cite{CL20} and quantized Thouless pumping~\cite{SCL1}. The modified CL with interacting particles have also been employed to study topological superconductors~\cite{CL21} and fractional topological insulators~\cite{CL9,CL22}.

Our manuscript is organized as follows. We start by introducing an experimentally realizable
quasi-1D ladder model of spin-$1/2$ fermions, which belongs to the
symmetry class CII of the periodic table~\cite{Tenfold1,Tenfold2} and corresponds to a spin-$1/2$ version of the CL. Our Floquet system is then
obtained by applying periodic quenches to the physical parameters
of the ladder. Thanks to the driving fields, rich Floquet topological
phases are found in this periodically quenched ladder model. A pair
of topological winding numbers are then introduced to characterize these
new phases, which can in principle take arbitrarily large even integer
values for the model we considered. Under the OBC, we further obtain
many quartets of Floquet topological zero and $\pi$ edge modes, whose
numbers are determined by the bulk topological invariants we introduced.
This establishes the bulk-edge correspondence of 1D Floquet systems
in the CII symmetry class. Finally, we propose to detect the Floquet
topological phases in our system by measuring the chiral displacements
of wavepackets. We show that from these displacements, a pair of dynamical
winding numbers can be constructed, which are equal to the topological
invariants of the system. In Sec. \ref{sec:Summary}, we summarize
our results and discuss potential future directions.

\section{The model\label{sec:Model}}

In this section, we introduce the model that will be explored in this
work. We will consider an extended version of the CL model with spin-$1/2$
fermions and spin-orbit couplings in the ladder. Recently, such kinds
of spin-$1/2$ CL (SCL) have been investigated in several studies~\cite{SCL1,SCL2,SCL3}. It
was found that the existence of spin degrees of freedom could not
only modify the symmetry classification of the CL (e.g., from BDI to CII), but also induce
new topological transport phenomena. However, in these studies only
a few topological phases of the SCL with small winding numbers were
identified, and the richness of topological states in the SCL has
not been uncovered. In this study, we will reveal the possible new
topological phases that can appear in the SCL with the help of Floquet
engineering.

\begin{figure}
\begin{centering}
\includegraphics[scale=0.295]{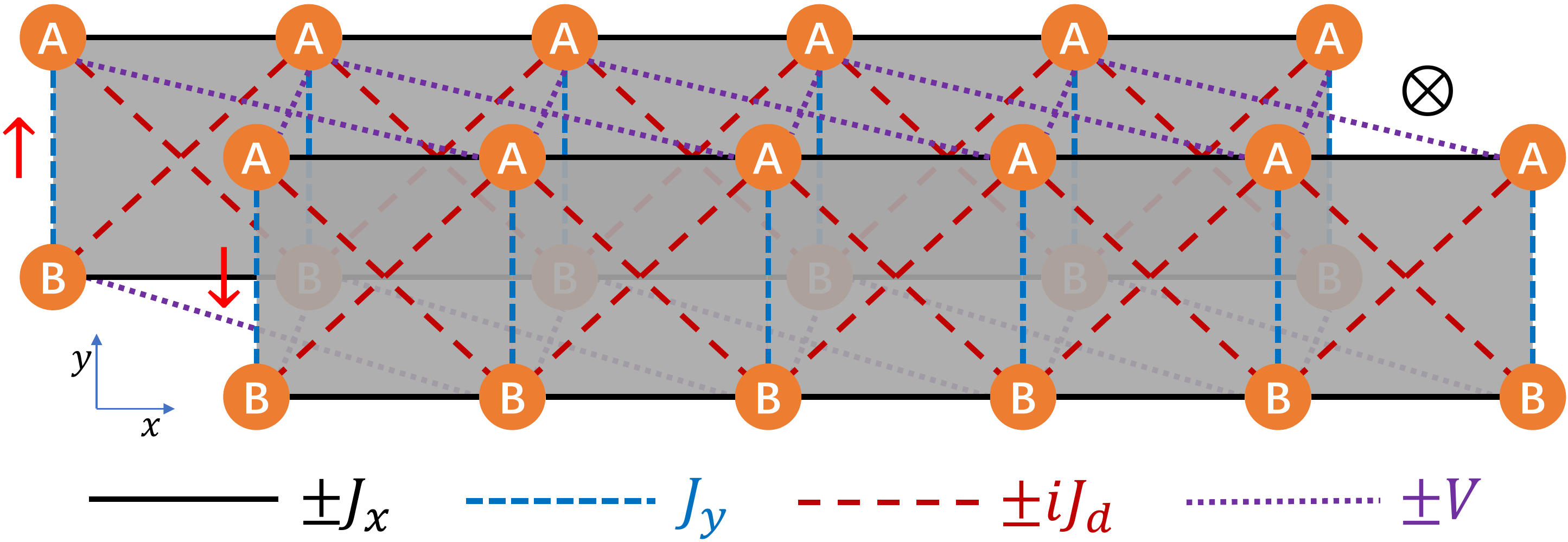}
\par\end{centering}
\caption{Schematic diagram of the spin-$1/2$ Creutz ladder. The backward (forward)
image corresponds to the spin $\uparrow$ ($\downarrow$) copy of
the ladder. Each unit cell contains two sublattices $A$ and $B$.
The magnetic flux ($\otimes$) is perpendicular to each plaquette
(shaded square). The nearest neighbor hopping amplitudes along $x$,
$y$ and diagonal directions of the ladder are denoted by $J_{x}$,
$J_{y}$ and $J_{d}$, respectively. The spin-orbit interaction $V$
couples particles with opposite spins in adjacent unit cells.\label{fig:SCL}}
\end{figure}

To do so, we first introduce our version of the SCL model, which is
schematically shown in Fig.~\ref{fig:SCL}. The model can be described
by the Hamiltonian
\begin{equation}
H_{0}=H_{1}+H_{2},\label{eq:H0}
\end{equation}
where the two components $H_{1}$ and $H_{2}$ are given by:
\begin{alignat}{1}
H_{1}= & \sum_{n}J_{x}(|n\rangle\langle n+1|+{\rm H.c.})\sigma_{0}\otimes\tau_{z}\nonumber \\
- & \sum_{n}iV(|n\rangle\langle n+1|-{\rm H.c.})\sigma_{y}\otimes\tau_{0},\label{eq:H1}\\
H_{2}= & \sum_{n}J_{y}|n\rangle\langle n|\sigma_{0}\otimes\tau_{x}\nonumber \\
+ & \sum_{n}iJ_{d}(|n\rangle\langle n+1|-{\rm H.c.})\sigma_{z}\otimes\tau_{x}.\label{eq:H2}
\end{alignat}
Here $n$ is the unit cell index. Each unit cell of the ladder
contains two sublattices $A$ and $B$. $J_{x}$ and $J_{y}$ are
intercell and intracell hopping amplitudes along the $x$ and $y$
directions of the ladder. $J_{d}$ is the diagonal hopping amplitude
between different sublattices in adjacent unit cells, with the prefactor
$i$ originated from the magnetic flux perpendicular to the plaquettes
of the ladder. $V$ is the strength of spin-orbit coupling between
electrons of different spins in the same sublattice of different unit cells. The Pauli matrix
$\sigma_{x,y,z}$ and $\tau_{x,y,z}$ act on the spin-$1/2$ and sublattice
degrees of freedom, respectively, and $\sigma_{0},\tau_{0}$ both
represent the $2\times 2$ identity matrix. In the lattice representation,
it is straightforward to see that our SCL Hamiltonian $H_0$ belongs to the symmetry
class CII of topological insulators. It possesses the time reversal
symmetry (TRS) $\mathfrak{T}=i\sigma_{y}\otimes\tau_{0}K$ with $\mathfrak{T}^{2}=-1$,
particle hole symmetry (PHS) $\mathfrak{C}=\sigma_{x}\otimes\tau_{y}K$
with $\mathfrak{C}^{2}=-1$, and chiral symmetry (CS) ${\cal S}=\mathfrak{T}\mathfrak{C}=-\sigma_{z}\otimes\tau_{y}$
with ${\cal S}^{2}=1$, in the sense that $\mathfrak{T}H_{0}\mathfrak{T}^{-1}=H_{0}$,
$\mathfrak{C}H_{0}\mathfrak{C}^{-1}=-H_{0}$ and ${\cal S}H_{0}{\cal S}=-H_{0}$.
The topological phases of $H_0$ are thus characterized by a winding
number $w$, which can only take even integer values $(2\mathbb{Z})$
due to the symmetry constraints~\cite{Tenfold1}. Furthermore, under the OBC, fourfold
degenerate edge modes are expected to appear when the SCL is in a
topologically nontrivial phase. The exact number of these edge
modes is determined by the bulk topological invariant $w$.

We now incorporate our Floquet engineering approach, with the purpose
of realizing new topological phases with large winding numbers and
many degenerate edge modes in the SCL. We consider a simple scheme,
in which the SCL is piecewise quenched within each driving period.
Such a periodically quenched SCL (PQSCL) is described by the Hamiltonian:
\begin{equation}
H(t)=\begin{cases}
H_{1} & t\in[jT,jT+T/2)\\
H_{2} & t\in[jT+T/2,jT+T)
\end{cases},\label{eq:Ht}
\end{equation}
where $j\in\mathbb{Z}$, $t$ is time and $T$ is the driving period.
Within each driving period, the Hamiltonians $H_{1}$ and $H_{2}$
are given by Eqs.~(\ref{eq:H1}) and (\ref{eq:H2}). The Floquet operator
of the PQSCL, which governs its dynamics over a complete driving period
$T$~[e.g., from $t=jT+0^-\rightarrow(j+1)T+0^-$], is then given by
\begin{equation}
U\equiv\mathsf{T}e^{-\frac{i}{\hbar}\int_{0}^{T}H(t)dt}=e^{-\frac{i}{2}H_{2}}e^{-\frac{i}{2}H_{1}},\label{eq:U}
\end{equation}
where $\mathsf{T}$ is the time ordering operator. In the second
equality, we have set $\hbar=T=1$ and choose the unit of energy to
be $\hbar/T$. Under the OBC, the quasienergy spectrum $\varepsilon$
and Floquet eigenstates of the PQSCL are obtained by solving
the eigenvalue equation $U|\psi\rangle=e^{-i\varepsilon}|\psi\rangle$.
Under the PBC, we can perform a Fourier transform and obtain the Floquet
operator in the momentum representation as
\begin{equation}
U(k)=e^{-ih_{2}(k)}e^{-ih_{1}(k)},\label{eq:Uk}
\end{equation}
where 
\begin{alignat}{1}
h_{1}(k) & = J_{x}\cos k\sigma_{0}\otimes\tau_{z}+V\sin k\sigma_{y}\otimes\tau_{0},\label{eq:h1k}\\
h_{2}(k) & = \frac{J_{y}}{2}\sigma_{0}\otimes\tau_{x}-J_{d}\sin k\sigma_{z}\otimes\tau_{y},\label{eq:h2k}
\end{alignat}
and $k\in[-\pi,\pi)$ is the quasimomentum. In the basis of Floquet-Bloch
eigenstates, $U(k)$ can also be expressed as 
\begin{equation}
U(k)=\sum_{\ell=1,2}\sum_{\eta=\pm}e^{-i\varepsilon_{\ell}^{\eta}(k)}|\varepsilon_{\ell}^{\eta}(k)\rangle\langle\varepsilon_{\ell}^{\eta}(k)|,\label{eq:UkEB}
\end{equation}
where $\varepsilon_{\ell}^{\pm}(k)=\pm\varepsilon_{\ell}(k)$ is
the quasienergy dispersion, and $|\varepsilon_{\ell}^{\pm}(k)\rangle$
are the corresponding Floquet eigenstates. The indices $\ell=1,2$ count the two Floquet bands whose quasienergies are in the range $(0,\pi)$.
Though the Floquet operator $U(k)$ describes a four-band model, its analytical diagonalization is complicated due to the presence of spin-orbit coupling. Instead, the quasienergy spectrum and Floquet eigenstates of $U(k)$ can be easily found by numerical calculations.
In the next section, we will unravel the richness of the bulk 
spectrum and topological properties of the PQSCL described by $U(k)$.

\section{Bulk topological properties\label{sec:Bulk}}

In this section, we focus on the characterization of bulk Floquet
topological phases of the PQSCL.
It is clear that the Floquet operator $U(k)$ in Eq.~(\ref{eq:Uk}) does not explicitly possess the TRS, PHS and CS of the non-driven SCL Hamiltonian $H_0$.
The classification of Floquet operators like $U(k)$ in one-dimension then rely on the introduction of a pair of symmetric time frames~\cite{AsbothSTF}. Upon similarity transformations, the bulk Floquet operator $U(k)$ in Eq.~(\ref{eq:Uk}) can be expressed in these time frames as
\begin{alignat}{1}
U_{1}(k)= & e^{-\frac{i}{2}h_{1}(k)}e^{-ih_{2}(k)}e^{-\frac{i}{2}h_{1}(k)},\label{eq:U1k}\\
U_{2}(k)= & e^{-\frac{i}{2}h_{2}(k)}e^{-ih_{1}(k)}e^{-\frac{i}{2}h_{2}(k)},\label{eq:U2k}
\end{alignat}
where $h_{1}(k)$ and $h_{2}(k)$ are given by Eqs.~(\ref{eq:h1k}) and (\ref{eq:h2k}). It is clear
that the Floquet operators $U(k)$, $U_{1}(k)$ and $U_{2}(k)$ share
the same quasienergy spectrum. In the meantime, the Floquet operators
$U_{1}(k)$ and $U_{2}(k)$ both possess the TRS ${\cal T}=i\sigma_{y}\otimes\tau_{0}$,
PHS ${\cal C}=\sigma_{x}\otimes\tau_{y}$ and CS ${\cal S}={\cal TC}^{*}=-\sigma_{z}\otimes\sigma_{y}$,
in the sense that
\begin{alignat}{1}
{\cal T}U_{\alpha}^{*}(k){\cal T}^{\dagger}& = U_{\alpha}^{\dagger}(-k),\nonumber \\
{\cal C}U_{\alpha}^{*}(k){\cal C}^{\dagger}& = U_{\alpha}(-k),\label{eq:Symmetry}\\
{\cal S}U_{\alpha}(k){\cal S}& = U_{\alpha}^{\dagger}(k),\nonumber 
\end{alignat}
for $\alpha=1,2$. Moreover, since we have ${\cal T}{\cal T}^{*}=-1$,
${\cal C}{\cal C}^{*}=-1$ and ${\cal S}^{2}=1$, the Floquet operators
$U_{1}(k)$ and $U_{2}(k)$ belong to the symmetry class CII according
to the periodic table of Floquet topological insulators~\cite{SymCls2}. Therefore,
the topological phases of Floquet operator $U(k)$ are characterized
by a pair of winding numbers, which can only take even integer values
($2\mathbb{Z}\times2\mathbb{Z}$). 

To obtain these topological invariants, we first introduce winding
numbers $w_{1}$ and $w_{2}$ for the Floquet operators $U_{1}(k)$
and $U_{2}(k)$ in symmetric time frames, i.e.,
\begin{equation}
w_{\alpha}=\int_{-\pi}^{\pi}\frac{dk}{4\pi i}{\rm Tr}[{\cal S}{\cal Q}_{\alpha}(k)\partial_{k}{\cal Q}_{\alpha}(k)]\label{eq:W12}
\end{equation}
for $\alpha=1,2$. Here ${\cal S}$ is the chiral symmetry operator.
The ${\cal Q}$-matrix ${\cal Q}_{\alpha}(k)$ is defined as
\begin{equation}
{\cal Q}_{\alpha}(k)\equiv\sum_{\ell=1,2}\sum_{\eta=\pm}\eta|\varepsilon_{\alpha\ell}^{\eta}(k)\rangle\langle\varepsilon_{\alpha\ell}^{\eta}(k)|,\label{eq:Q12k}
\end{equation}
where $|\varepsilon_{\alpha\ell}^{\eta}(k)\rangle$ is the Floquet
eigenstate of $U_{\alpha}(k)$ with quasienergy $\varepsilon_{\ell}^{\eta}(k)=\eta\varepsilon_{\ell}(k)$,
and $\ell=1,2$ are the indices of the two Floquet bands, whose quasienergies
are in the range $(0,\pi)$. ${\cal Q}_{\alpha}(k)$ can thus be viewed
as a ``flat-band version'' of the Floquet effective Hamiltonian
of $U_{\alpha}(k)$, in the sense that all the positive (negative)
quasieneriges are set to $1$ ($-1$). It is clear that such a band
flattening procedure does not change the topological properties of
$U_{\alpha}(k)$, so long as its Floquet spectrum is gapped at the
quasienergies zero and $\pi$ during the flattening process.

In terms of the winding numbers $w_{1}$ and $w_{2}$, we can define the invariants that characterize the Floquet topological phases of the PQSCL. 
The construction of these invariants is inspired by the known characterization of chiral symmetric Floquet topological matter~\cite{AsbothSTF}. Since for a chiral symmetric Floquet operator, two topological spectral gaps are allowed at quasienergies $0$ and $\pi$, we need a pair of invariants to fully characterize its topological phases. As both $U_1(k)$ and $U_2(k)$ are chiral symmetric, an appropriate combination of their winding numbers should be able to produce the topological invariants of the system.
Following previous studies~\cite{AsbothSTF}, these invariants can be defined as
\begin{equation}
w_{0}=\frac{w_{1}+w_{2}}{2},\qquad w_{\pi}=\frac{w_{1}-w_{2}}{2}.\label{eq:W0P}
\end{equation}
According to the symmetry classification scheme of our Floquet system, we
have $(w_{0},w_{\pi})\in2\mathbb{Z}\times2\mathbb{Z}$, and their
values specify all possible Floquet topological phases that can appear
in the PQSCL described by the Floquet operator $U(k)$ in Eq.~(\ref{eq:Uk}). 

In the remaining part of this section, we verify \textcolor{red}{Eq.~(\ref{eq:W0P})}
for the PQSCL numerically, and then construct its Floquet topological phase
diagram. To do so, we first introduce a pair of gap characteristic
functions, defined as
\begin{alignat}{1}
\Delta_{0}\equiv & \min_{k\in[-\pi,\pi)}^{\ell\in\{1,2\}}|\cos[\varepsilon_{\ell}(k)]-1|,\label{eq:DEL0}\\
\Delta_{\pi}\equiv & \min_{k\in[-\pi,\pi)}^{\ell\in\{1,2\}}|\cos[\varepsilon_{\ell}(k)]+1|,\label{eq:DELP}
\end{alignat}
where $\varepsilon_{\ell}(k)$ is the quasienergy of the eigenstate
$|\varepsilon_{\alpha\ell}^{+}(k)\rangle$ of $U_{\alpha}(k)$ [see Eq.~(\ref{eq:Q12k})]. It
is clear that due to the CS and the $2\pi$-periodicity of the quasienergy,
the Floquet spectrum of PQSCL will become gapless at the quasienergy
zero ($\pi$) if $\Delta_{0}=0$ ($\Delta_{\pi}=0$). A topological
phase transition accompanied by the quantized change of winding number
$w_{0}$ ($w_{\pi}$) may occur when the gap function $\Delta_{0}$
($\Delta_{\pi}$) vanishes.

\begin{figure}
\begin{centering}
\includegraphics[scale=0.49]{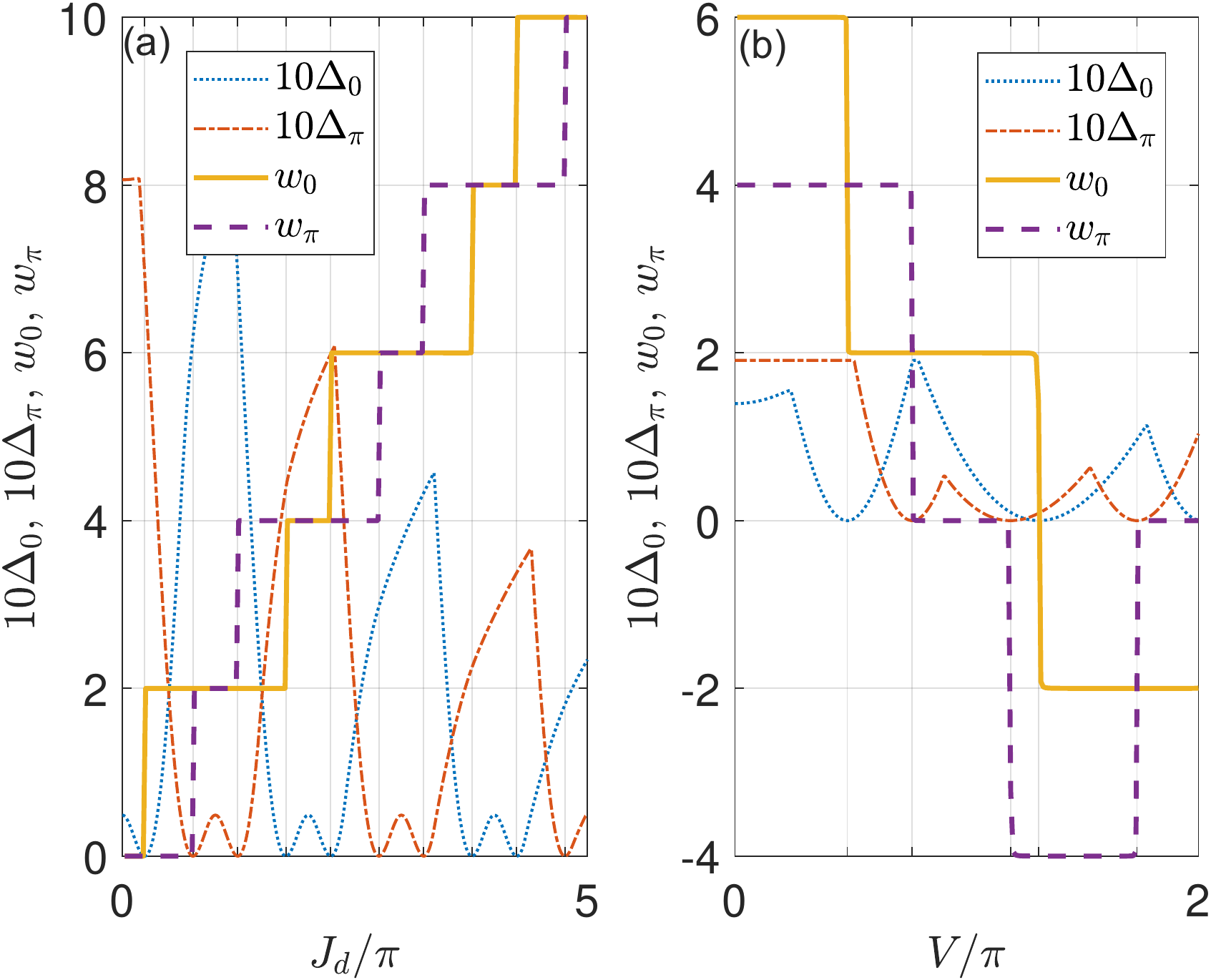}
\par\end{centering}
\caption{Gap functions $\Delta_{0}$ (blue dotted lines), $\Delta_{\pi}$ (red
dash-dotted lines), and winding numbers $w_{0}$ (yellow solid lines),
$w_{\pi}$ (purple dashed lines) versus the diagonal hopping amplitude
$J_{d}$ and spin-orbit coupling strength $V$ in panels (a) and (b),
respectively. The other system parameters are set as $J_{x}=0.5\pi$,
$J_{y}=0.6\pi$, and $V=0.2\pi$ ($J_{d}=2.5\pi$) for panel (a) {[}(b){]}.
The ticks along the horizontal axis correspond to the local minimum
of gap functions $(\Delta_{0},\Delta_{\pi})$, which are obtained
from Eqs.~(\ref{eq:DEL0}) and (\ref{eq:DELP}).\label{fig:GFWN}}
\end{figure}

In Figs.~\ref{fig:GFWN}(a) and (b), we present the (rescaled) gap
functions $\Delta_{0}$ (blue dotted lines), $\Delta_{\pi}$ (red
dash-dotted lines) and the winding numbers $w_{0}$ (yellow solid
lines), $w_{\pi}$ (purple dashed lines) with respect to the change
of diagonal hopping amplitude $J_{d}$ and spin-orbit coupling strength
$V$, respectively. In both panels, we observe that the winding number
$w_{0}$ or $w_{\pi}$ indeed shows a quantized jump everytime when $\Delta_{0}=0$
or $\Delta_{\pi}=0$, reflecting the existence of Floquet topological
phase transitions in the PQSCL. Furthermore, with the increase of
$J_{d}$, we obtain topological phases with larger and larger
winding numbers $(w_{0},w_{\pi})$. Such a trend will continue, and
we could in principle find Floquet topological phases in the PQSCL,
whose winding numbers $(w_{0},w_{\pi})$ could take arbitrarily large even
integers. To the best of our knowledge, this is the first proposal
of a topological insulator in the CII symmetry class with unlimited winding numbers
under the framework of Floquet engineering.

\begin{figure}
\begin{centering}
\includegraphics[scale=0.5]{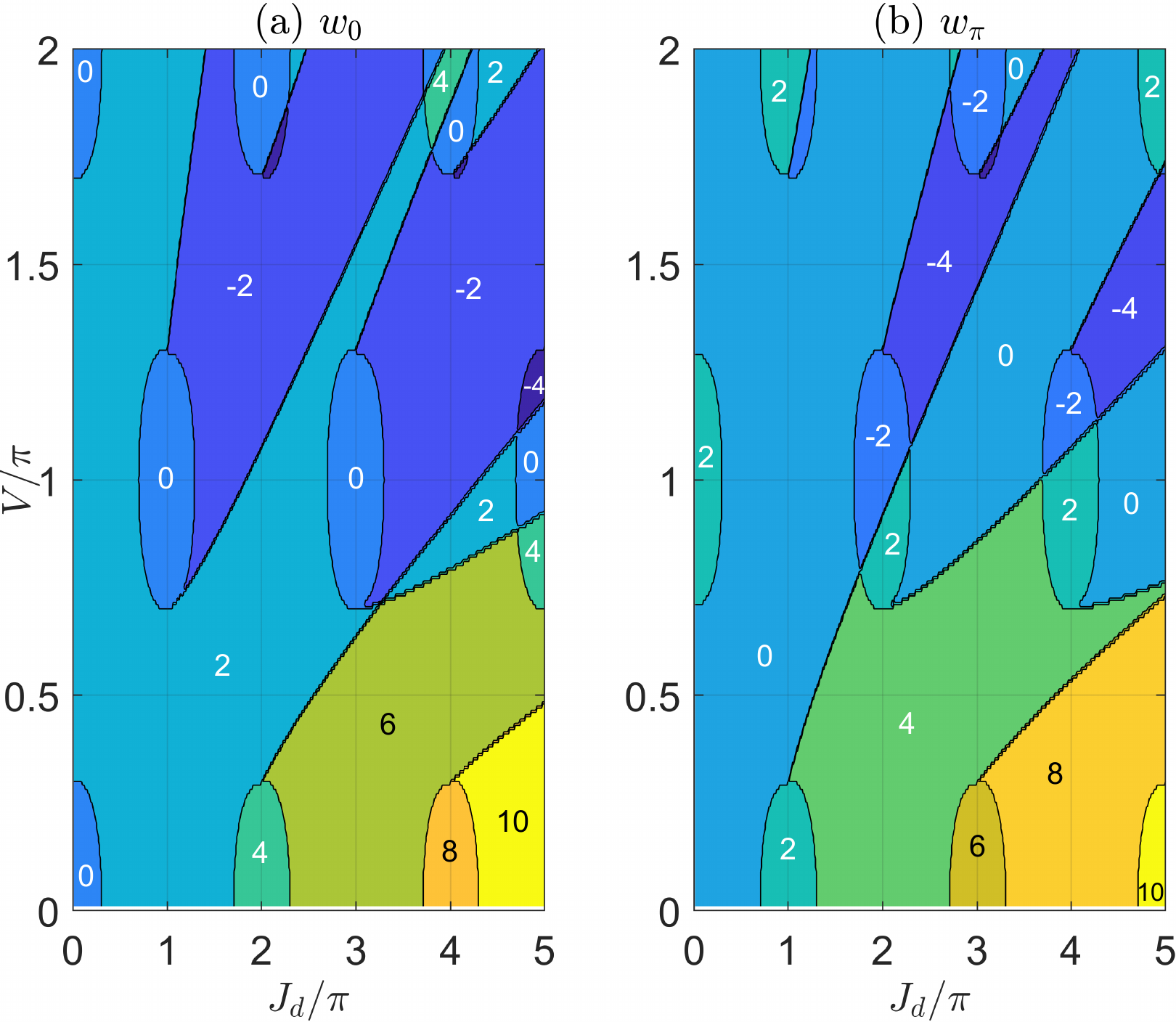}
\par\end{centering}
\caption{The topological invariants $w_{0}$ {[}panel (a){]} and $w_{\pi}$
{[}panel (b){]} of the PQSCL versus the diagonal hopping amplitude
$J_{d}$ and the spin-orbit coupling strength $V$. Other system parameters
are chosen as $(J_{x},J_{y})=(0.5\pi,0.6\pi)$. Each region with a
uniform color corresponds to a Floquet topological phase, with the
corresponding value of winding number $w_{0}$ or $w_{\pi}$ being
denoted explicitly and calculated by Eqs.~(\ref{eq:W12})-(\ref{eq:W0P}). The black lines separating different regions are the boundaries between different topological phases, obtained from Eqs.~(\ref{eq:DEL0}) and (\ref{eq:DELP}).\label{fig:PhsDiag}}
\end{figure}
To fully reveal the Floquet topological phases that can appear in
the PQSCL model, we present its topological phase diagram versus the
system parameters $(J_{d},V)$ in Fig.~\ref{fig:PhsDiag} for a typical
situation. The other system parameters are chosen as $J_{x}=0.5\pi$
and $J_{y}=0.6\pi$. In the phase diagram, each region with a uniform
color corresponds to a Floquet topological phase, whose winding numbers
$(w_{0},w_{\pi})$ are denoted explicitly in the region. Across each
line between two adjacent regions, the winding numbers $w_{0}$ or
$w_{\pi}$ is changed by a multiple of an even integer, corresponding to a Floquet
topological phase transition. 
Note in passing that by choosing isotropic hopping amplitudes along $x$ and $y$ directions of the ladder~(i.e., set $J_x=J_y=0.5\pi$ here), we will get a similar topological phase diagram compared with Fig.~\ref{fig:PhsDiag}, with only small deviations of the phase boundaries relative to the case of anisotropic hopping amplitudes.
Therefore, we conclude that the PQSCL indeed possesses
rich Floquet topological phases, with many of them being characterized by
large and even-integer topological winding numbers. The existence of
these nontrivial phases have direct implications to the edge states
and transport properties of the system, as will be explored in the
next two sections.

\section{Edge states and bulk-edge correspondence\label{sec:Edge}}
%In the previous section, we have shown that the PQSCL model belongs to the symmetry class CII, and each of its Floquet topological phases is characterized by a pair of even integer winding numbers $(w_{0},w_{\pi})\in2\mathbb{Z}\times2\mathbb{Z}$.
When the bulk Floquet spectrum of PQSCL is gapped at the quasienergies zero and $\pi$, we could further obtain zero and $\pi$ Floquet edge modes under the OBC, and their numbers $(n_{0},n_{\pi})$ should be related to the bulk topological invariants $(w_{0},w_{\pi})$ in Eq.~(\ref{eq:W0P}), i.e., $n_{0}=2|w_{0}|$ and $n_{\pi}=2|w_{\pi}|$. These relations describe the bulk-edge correspondence of 1D Floquet topological insulators in the CII class, as will be demonstrated numerically in this section.

More generally, one can also introduce the topological invariants
of the system directly under the OBC. To do so, we first define the
non-commutative winding number in the symmetric time frame $\alpha$
($=1,2$) under the OBC as~\cite{OBCWN1}
\begin{equation}
\widetilde{w}_{\alpha}\equiv\frac{1}{2N_{b}}{\rm Tr}_{b}({\cal S}{\cal Q}_{\alpha}[{\cal Q}_{\alpha},\hat{n}]).\label{eq:WTid12}
\end{equation}
Here ${\cal S}$ is the chiral symmetry operator, and $\hat{n}=\sum_{n=1}^{N}n|n\rangle\langle n|\sigma_{0}\otimes\tau_{0}$
is the position operator of unit cell. The total number of unit cells
of the ladder is $N=N_{b}+2N_{e}$, with $N_{b}$ and $N_{e}$ being
the number of cells in the bulk and edge intervals of the ladder,
respectively. In terms of the unit cell coordinate $n$, we have $n\in[N_{e}+1,N_{e}+N_{b}]$
for the bulk interval and $n\in[1,N_{e}]\cup[N-N_{e}+1,N]$ for the
left and right edge intervals. The trace ${\rm Tr}_{b}(\cdot)$ in Eq.~(\ref{eq:WTid12})
is taken only over the degrees of freedom in the bulk interval $n\in[N_{e}+1,N_{e}+N_{b}]$.
In numerical calculations, $N_{e}$ should be chosen large enough
in order to avoid boundary effects. The open boundary ${\cal Q}$-matrix~\cite{OBCWN1}
${\cal Q}_{\alpha}$ in Eq.~(\ref{eq:WTid12}) is given by
\begin{equation}
{\cal Q}_{\alpha}\equiv\sum_{n=1}^{2N}\sum_{\eta=\pm}\eta|\varepsilon_{\alpha n}^{\eta}\rangle\langle\varepsilon_{\alpha n}^{\eta}|\label{eq:Q12OBC}
\end{equation}
for the two time frames $\alpha=1,2$, where the Floquet eigenstates $\{|\varepsilon_{\alpha n}^{\eta}\rangle\}$
satisfy the eigenvalue equation $U_{\alpha}|\varepsilon_{\alpha n}^{\eta}\rangle=e^{-i\varepsilon_{n}^{\eta}}|\varepsilon_{\alpha n}^{\eta}\rangle=e^{-i\eta\varepsilon_{n}}|\varepsilon_{\alpha n}^{\eta}\rangle$
for all the quasienergy eigenvalues $\{\varepsilon_{1},...,\varepsilon_{2N}\}\in(0,\pi)$.
Physically, ${\cal Q}_{\alpha}$ can be understood as a flattened
effective Floquet Hamiltonian of the system, whose positive and negative
quasienergies are set to $1$ and $-1$, respectively. With $(\widetilde{w}_{1},\widetilde{w}_{2})$
defined in Eq.~(\ref{eq:WTid12}), we can construct a pair of topological invariants for
the system under the OBC as
\begin{equation}
\widetilde{w}_{0}=\frac{\widetilde{w}_{1}+\widetilde{w}_{2}}{2},\qquad\widetilde{w}_{\pi}=\frac{\widetilde{w}_{1}-\widetilde{w}_{2}}{2}.\label{eq:WTid0P}
\end{equation}
Since in a given time frame $\alpha$ ($=1,2$), early studies~\cite{OBCWN2} have
established that $w_{\alpha}=\widetilde{w}_{\alpha}$, we could finally
state the bulk-edge correspondence of our PQSCL model as:
\begin{equation}
(n_{0},n_{\pi})=(2|w_{0}|,2|w_{\pi}|)=(2|\widetilde{w}_{0}|,2|\widetilde{w}_{\pi}|).\label{eq:BBC}
\end{equation}
Note that the second equality holds even when the system possesses
disorder, assuming that the disorder does not break the TRS, PHS and
CS of the model. Furthermore, since $w_{0}$ and $w_{\pi}$ are both
even integers, the values of $n_{0}$ and $n_{\pi}$ must be integer
multiples of four. This implies that under the OBC, the Floquet zero and $\pi$ edge
modes always appear as quartets in the PQSCL, with their fourfold degeneracy being
protected by the symmetries of the system.

\begin{figure}
\begin{centering}
\includegraphics[scale=0.49]{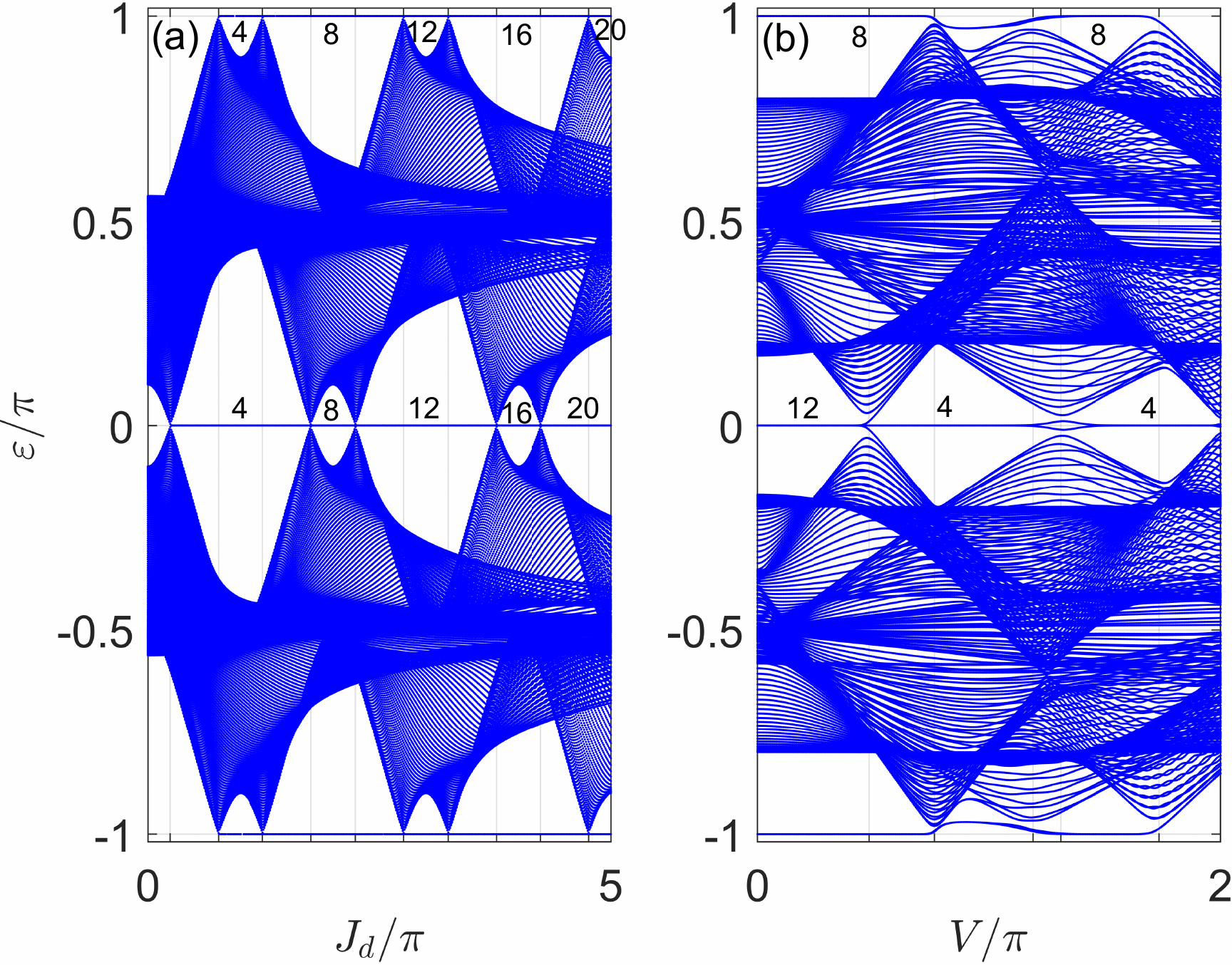}
\par\end{centering}
\caption{The quasienergy spectrum of PQSCL under the OBC versus the diagonal
hopping amplitude $J_{d}$ and the spin-orbit coupling strength $V$
in panels (a) and (b), respectively. The hopping amplitudes along
$x$ and $y$ directions are chosen to be $(J_{x},J_{y})=(0.5\pi,0.6\pi)$
for both panels, and the total number of unit cells is $N=200$. Other
system parameters are set at $V=0.2\pi$ for panel (a) and $J_{d}=2.5\pi$
for panel (b). The ticks along the horizontal axis correspond to the
local minimum of bulk gap functions $(\Delta_{0},\Delta_{\pi})$,
which are obtained from Eqs.~(\ref{eq:DEL0}) and (\ref{eq:DELP}).
In both panels, the numbers appearing around $\varepsilon=0$ and $\pi$ are the numbers of degenerate topological edge modes $n_0$ and $n_{\pi}$ at these quasienergies.\label{fig:Spectrum_OBC}}
\end{figure}

To demonstrate the edge states and bulk-edge correspondence of the
PQSCL model, we present its quasienergy spectrum under the OBC for
two typical situations. In Fig.~\ref{fig:Spectrum_OBC}(a), we show
the Floquet spectrum $\varepsilon$ of the system versus the diagonal
hopping amplitude $J_{d}$. The other system parameters are chosen
as $J_{x}=0.5\pi$, $J_{y}=0.6\pi$, $V=0.2\pi$, and the ladder contains
a number of $N=200$ unit cells. From Fig. \ref{fig:Spectrum_OBC}(a),
we see that across each topological phase transition shown in the
phase diagram Fig.~\ref{fig:PhsDiag}, the spectrum gap closes and reopens at the
quasienergy zero or $\pi$, accompanied by the emergence of new Floquet
zero or $\pi$ edge modes. The numbers of these edge modes $n_{0}$
and $n_{\pi}$ are denoted explicitly for each phase in the figure,
and their values are consistent with the bulk-edge correspondence
as described by Eq.~(\ref{eq:BBC}). Notably, with the increase of
$J_{d}$, the system undergoes a series of topological phase
transitions, with more and more edge modes appear at both zero and
$\pi$ quasienergies. Therefore, we can in principle obtain arbitrarily
many quartets of zero and $\pi$ Floquet edge modes by simply tuning
the diagonal hopping amplitude $J_{d}$ of the ladder. The observation
of these edge modes in atom-optical quantum simulators could not only
verify the topological nature of the PQSCL model, but also demonstrate
the power of Floquet engineering in the realization of new phases
with large topological invariants.

In Fig.~\ref{fig:Spectrum_OBC}(b), we show the Floquet spectrum $\varepsilon$
of the PQSCL versus the spin-orbit coupling strength $V$. The other
system parameters are fixed at $J_{x}=0.5\pi$, $J_{y}=0.6\pi$, $J_{d}=2.5\pi$,
and the number of unit cells $N=200$. Again, we observe Floquet zero
and $\pi$ edge modes in each of the topologically nontrivial phases, with
their numbers $(n_{0},n_{\pi})$ being related to the winding numbers
$(w_{0},w_{\pi})$ through the bulk-edge relation Eq.~(\ref{eq:BBC}).
Across each topological phase transition (i.e., gap closing and reopening) point, the
number of edge modes at quasienergies zero or $\pi$ also appear or
disappear as quartets, as predicted by our theory. Note that the numbers of edge
states $(n_{0},n_{\pi})$ change non-monotonically with the
increase of $V$, with no tendency of raising as compared with the
case of $(n_{0},n_{\pi})$ versus $J_{d}$.

To sum up, we have established the relation between the bulk topological
invariants $(w_{0},w_{\pi})$ of the PQSCL and the number of its fourfold
degenerate edge modes $(n_{0},n_{\pi})$ at both quasieneriges zero
and $\pi$ under the OBC, as given by Eq.~(\ref{eq:BBC}). In the
next section, we further construct a dynamical observable, which could
help us to detect the topological invariants and topological phase transitions of
the PQSCL model.

\section{Dynamical probes\label{sec:Dynamics}}
The mean chiral displacement (MCD) is an observable, which could directly
produce the topological winding numbers of a system with chiral symmetry
in its dynamical evolution~\cite{MCD1,MCD2,MCD3,MCD4,MCD5}. It is obtained by measuring the long-time
average of chiral displacement operator ${\cal S}\hat{n}$, where
${\cal S}$ is a unitary and Hermitian operator that defines the chiral
symmetry of the system, and $\hat{n}$ is the unit cell position operator.
The MCD was first proposed for 1D non-driven systems in the symmetry
classes AIII and BDI~\cite{MCD1}, and later extended to Floquet systems~\cite{LargeTI3,MCD2}, non-Hermitian systems~\cite{NHFTP2,NHFTP3} and systems in two spatial dimensions~\cite{HOFTP1}. In this work, we further extend
the MCD to 1D Floquet systems in the symmetry class CII, and employ
it to probe the topological phases of the PQSCL dynamically. 

We define the MCD as the long-time average of the chiral displacement
operator ${\cal S}\hat{n}$ in a given symmetric time frame. For a
1D Floquet system in the symmetry class CII, it can be expressed as
\begin{equation}
c_{\alpha}=\lim_{M\rightarrow\infty}\frac{1}{M}\sum_{m=1}^{M}\langle{\cal S}U_{\alpha}^{m}\hat{n}U_{\alpha}^{m}\rangle_{0},\label{eq:C12OBC}
\end{equation}
where $\alpha=1,2$ is the index of the symmetric time frame, $U_{\alpha}$
is the corresponding Floquet operator, $M$ is the total number of
driving periods (with the driving period $T=1$), ${\cal S}$ is the chiral symmetry
operator and $\hat{n}$ is the unit-cell position operator. $\langle\cdots\rangle_{0}\equiv{\rm Tr}(\rho_{0}\cdots)$
performs an average over the initial state $\rho_{0}$. For our PQSCL,
$\rho_{0}$ can be chosen as an incoherent summation of eigenmodes
occupying each spin and sublattice degrees of freedom in the central unit cell
($n=0$) of the lattice, i.e.,
\begin{equation}
\rho_{0}=\sum_{\sigma=\uparrow,\downarrow}\sum_{s=A,B}|0\sigma s\rangle\langle0\sigma s|.\label{eq:Rho0}
\end{equation}
Note here that the initial state $\rho_{0}$ is a mixed state. In
experiments, one can simply execute the dynamics for each component state
$|0\sigma s\rangle\langle0\sigma s|$ of $\rho_{0}$ separately, and
then sum up their contributions to $c_{\alpha}$. Taking the periodic
boundary condition, we can express $c_{\alpha}$ in momentum representation
as
\begin{equation}
c_{\alpha}=\lim_{M\rightarrow\infty}\frac{1}{M}\sum_{m=1}^{M}\int_{-\pi}^{\pi}\frac{dk}{2\pi}{\rm Tr}[{\cal S}U_{\alpha}^{m}(k)i\partial_{k}U_{\alpha}^{m}(k)],\label{eq:C12}
\end{equation}
where $k\in[-\pi,\pi)$ is the quasimomentum, and $\alpha=1,2$ are the indices of two time frames. 
For our PQSCL, $U_1(k)$ and $U_2(k)$ are given by Eqs.~(\ref{eq:U1k}) and (\ref{eq:U2k}), respectively.
With the help of $c_{1}$ and $c_{2}$, we can construct a pair of dynamical winding
numbers $c_{0}$ and $c_{\pi}$, defined as
\begin{equation}
c_{0}=\frac{c_{1}+c_{2}}{2},\qquad c_{\pi}=\frac{c_{1}-c_{2}}{2}.\label{eq:C0P}
\end{equation}
Then it can be shown that the dynamical winding numbers $(c_{0},c_{\pi})$
are equal to the topological invariants $(w_{0},w_{\pi})$ of the
PQSCL (see Appendix \ref{sec:AppA} for a proof), i.e.,
\begin{equation}
w_{0}=c_{0},\qquad w_{\pi}=c_{\pi}.\label{eq:WC0P}
\end{equation}
The relations from Eqs.~(\ref{eq:C12}) to (\ref{eq:WC0P}) establish the dynamical characterization of topological phases for 1D Floquet systems in the CII symmetry class, and provide a way for the detection of their topological invariants in quantum simulators like cold atoms and photonic setups.

\begin{figure}
\begin{centering}
\includegraphics[scale=0.5]{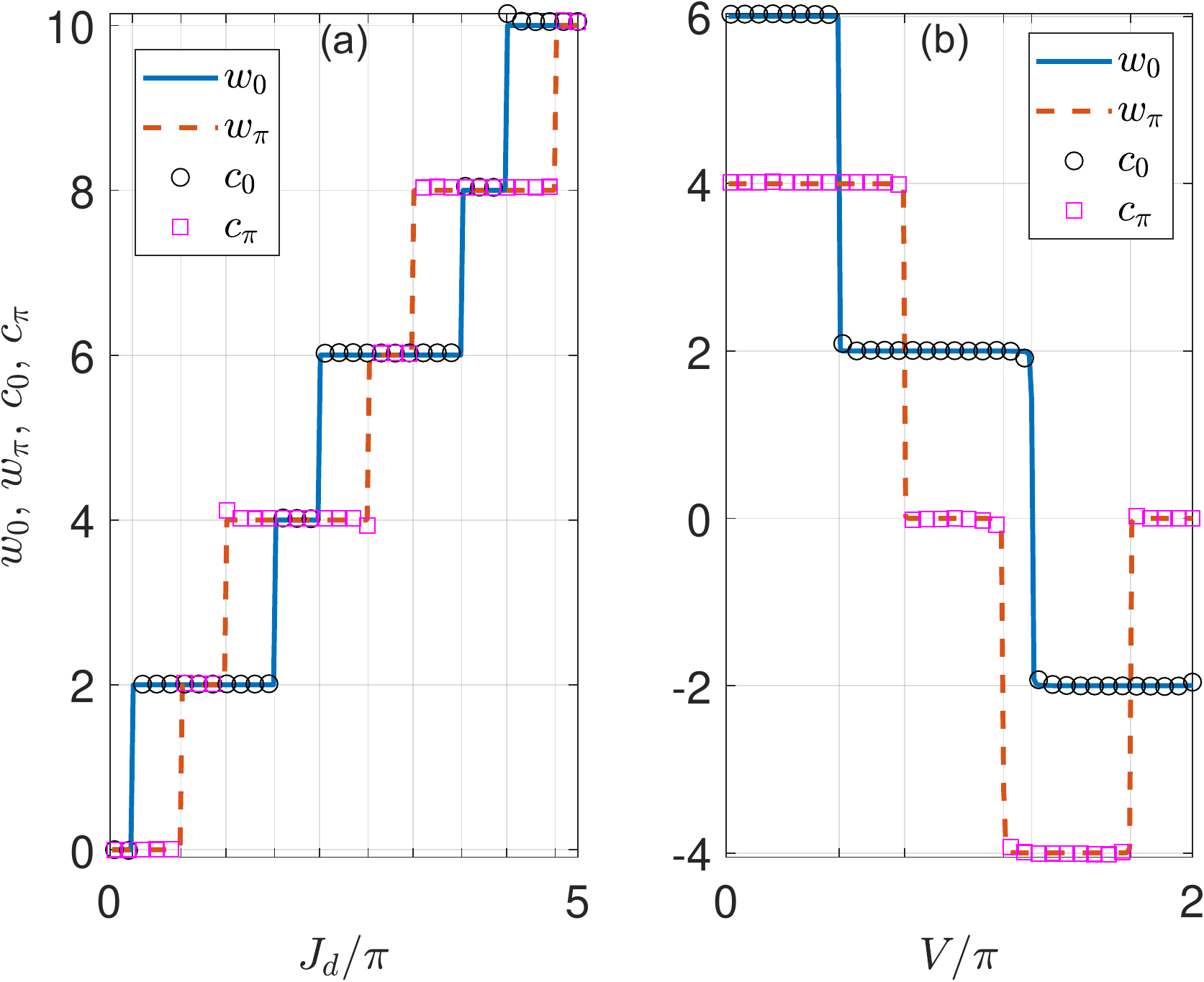}
\par\end{centering}
\caption{The topological invariants $w_{0}$ (blue solid lines), $w_{\pi}$
(red dashed lines), and dynamical winding numbers $c_{0}$ (black
circles), $c_{\pi}$ (magenta squares) versus the diagonal hopping
amplitude $J_{d}$ and spin-orbit coupling strength $V$ in panels
(a) and (b), respectively. The system parameters are chosen as $(J_{x},J_{y},V)=(0.5\pi,0.6\pi,0.2\pi)$
for panel (a) and $(J_{x},J_{y},J_{d})=(0.5\pi,0.6\pi,2.5\pi)$ for
panel (b). In both panels, $(w_{0},w_{\pi})$ are calculated according
to Eqs.~(\ref{eq:W12}) and (\ref{eq:W0P}), whereas $(c_{0},c_{\pi})$
are computed numerically following Eqs.~(\ref{eq:C12}) and (\ref{eq:C0P}),
with the total number of driving periods being $M=100$. The ticks along
the horizontal axis correspond to the local minimum of bulk gap functions
$(\Delta_{0},\Delta_{\pi})$, which are obtained from Eqs.~(\ref{eq:DEL0})
and (\ref{eq:DELP}).\label{fig:MCD}}
\end{figure}

In Figs.~\ref{fig:MCD}(a) and \ref{fig:MCD}(b), we numerically demonstrate the
relations in Eq.~(\ref{eq:WC0P}) for the PQSCL model in two typical situations.
In Fig.~\ref{fig:MCD}(a), we present the values of $w_{0}$ (blue
solid line), $w_{\pi}$ (red dashed line), $c_{0}$ (black circles)
and $c_{\pi}$ (magenta squares) versus the diagonal hopping amplitude $J_d$
of the PQSCL. Other system parameters are chosen to be $J_{x}=0.5\pi$,
$J_{y}=0.6\pi$, $V=0.2\pi$, and the MCDs are
averaged over $M=100$ driving periods to obtain the dynamical winding numbers. We observe that the behavior of dynamical winding numbers $(c_{0},c_{\pi})$ follow closely with
the theoretical predictions of topological invariants $(w_{0},w_{\pi})$,
as presented in Eq.~(\ref{eq:WC0P}). Notably, across each topological phase
transition point, the dynamical winding numbers $(c_{0},c_{\pi})$
show the same quantized jumps as $(w_{0},w_{\pi})$. Therefore, $(c_{0},c_{\pi})$
provide nice dynamical probes to the topological invariants and Floquet
topological phase transitions for our PQSCL model. In Fig.~\ref{fig:MCD}(b),
we further show the topological invariants $w_{0}$ (blue solid line),
$w_{\pi}$ (red dashed line) and dynamical winding numbers $c_{0}$
(black circles) and $c_{\pi}$ (magenta squares) versus the spin-orbit coupling
strength $V$ of the PQSCL model. The other system parameters are
fixed at $J_{x}=0.5\pi$, $J_{y}=0.6\pi$, $J_{d}=2.5\pi$ and $(c_{0},c_{\pi})$
are obtained by averaging the MCDs over $M=100$ driving periods. Again, we observe that
the values of $(w_{0},w_{\pi})$ and $(c_{0},c_{\pi})$ are consistent
with each other in each of the Floquet topological phases, and $(c_{0},c_{\pi})$
possess the same quantized jumps as $(w_{0},w_{\pi})$ across each
topological phase transition point. These observations further verify
the universality and correctness of Eq.~(\ref{eq:WC0P}) in characterizing and detecting
Floquet topological phases of the PQSCL. Note in passing that
the small deviations of $(c_{0},c_{\pi})$ from perfect quantization
in each topological phase is a finite-time effect, which can be suppressed
by increasing the total number of driving periods $M$ adopted in the average
of Eq.~(\ref{eq:C12}).

Practically, we have checked that the quantization of $(c_{0},c_{\pi})$
is already good enough for the number of evolution periods to be as
small as $M=15$. This should be already within reach in experimental
platforms like photonic and cold atom systems, where the MCDs have already
been measured for other lattice models~\cite{MCD1,MCD3,MCD4,MCD5}. 
In a photonic setup, the MCDs can be extracted from the Zak phases of the photonic quantum walk of twisted photons~\cite{MCD1}. In a cold atom setup, the MCDs may be obtained from the absorption images of the evolution of a wavepacket~\cite{MCD3}, which is initially prepared at the center of the lattice and then subjected to periodically quenched lattice parameters.
Combining these facts with
the existing realizations of the Creutz ladder model~\cite{CL14,CL15,CL16,CL17}, we expect that
our PQSCL and its topological properties should also be realizable
and detectable in similar experimental setups.

\section{Summary\label{sec:Summary}}

In this work, we proposed a periodically quenched spin-$1/2$ Creutz ladder
model, and investigated its Floquet topological phases. The model
belongs to the symmetry class CII in the periodic table of topological
matter. Its phases are characterized by a pair of winding numbers
$(w_{0},w_{\pi})$, which can only take even integer values ($2\mathbb{Z}\times2\mathbb{Z}$).
We established the phase diagram of the model and found rich Floquet
topological phases in the CII class, which possess large and even integer topological winding
numbers. Under the open boundary condition, we further obtained multiple
quartets of Floquet topological edge modes at both zero and $\pi$
quasienergies, with their numbers being determined by the bulk topological
invariants $(w_{0},w_{\pi})$. Finally, we showed that the topological
phases of the system can be probed by measuring a generalized mean
chiral displacement in two chiral symmetric time frames, and established the relation between the resulting dynamical winding numbers and the topological invariants of the system. Our discoveries thus introduced a new member to the zoo of Floquet topological phases,
which is featured by even integer topological invariants, fourfold degenerate edge modes, and also within reach in various atom-optical quantum simulators.

In recent years, the study of Floquet topological matter has been
extended to non-Hermitian domain, where the interplay between driving
fields and gain/loss or nonreciprocal effects could lead to unique
Floquet non-Hermitian topological phases with large topological invariants~\cite{NHFTP1,NHFTP2,NHFTP3,NHFTP4}.
It is expected that by adding non-Hermitian effects to our driven
Creutz ladder, Floquet phases with quadruplex edge modes could also
be induced even in the non-Hermitian regime, and the full characterization
of these new phases is an interesting direction to explore in future
studies. On application side, it is known that Majorana edge modes
could appear in the Creutz ladder with superconducting pairings. Therefore,
by adding superconducting pairing terms to our driven Creutz ladder,
we might be able to obtain many quartets of Floquet Majorana zero
and $\pi$ edge modes. Whether the fourfold degeneracy of these Majorana
modes could give more room for the braiding operations in the recently
proposed Floquet topological quantum computing~\cite{FloCompt1} is also an interesting
topic to explore in the future.

\section*{Acknowledgement}
L.Z. acknowledges Chushun Tian for helpful comments. This work is supported by the National Natural Science Foundation of China (Grant No.~11905211), the China Postdoctoral Science Foundation (Grant No.~2019M662444), the Fundamental Research Funds for the Central Universities (Grant No.~841912009), the Young Talents Project at Ocean University of China (Grant No.~861801013196), and the Applied Research Project of Postdoctoral Fellows in Qingdao (Grant No.~861905040009).

\appendix

\section{Relation between dynamical and topological winding numbers\label{sec:AppA}}
In this appendix, we prove that the dynamical winding numbers $(c_{0},c_{\pi})$
are equal to the topological invariants ($w_{0},w_{\pi}$) of the
PQSCL, as given by Eq.~(\ref{eq:WC0P}) in the main text. We begin
with the topological winding number $w_{\alpha}$ in the symmetric
time frame $\alpha$ ($=1,2$). In the momentum representation, with
the help of Eq.~(\ref{eq:W12}) and the ${\cal Q}$-matrix expression
in Eq.~(\ref{eq:Q12k}), we can express $w_{\alpha}$ as
\begin{alignat}{1}
w_{\alpha}= & \int_{-\pi}^{\pi}\frac{dk}{4\pi i}\sum_{\ell,\ell'=1,2}\sum_{\eta,\eta'=\pm}\eta\eta'\nonumber \\
\times & {\rm Tr}\{{\cal S}|\varepsilon_{\alpha\ell}^{\eta}(k)\rangle\langle\varepsilon_{\alpha\ell}^{\eta}(k)|\partial_{k}[|\varepsilon_{\alpha\ell'}^{\eta'}(k)\rangle\langle\varepsilon_{\alpha\ell'}^{\eta'}(k)|]\},\label{eq:WAapp1}
\end{alignat}
where $\eta=\pm$, and $\ell=1,2$ are the indices of the two Floquet
bands with quasienergies in the range $(0,\pi)$. Working out
the trace explicitly in the quasienergy eigenbasis $\{|\varepsilon_{\alpha\ell}^{\eta}\rangle\}$
of the Floquet operator $U_{\alpha}(k)$ in time frame $\alpha$,
we can further express $w_{\alpha}$ as:
\begin{alignat}{1}
w_{\alpha}= & \int_{-\pi}^{\pi}\frac{dk}{2\pi}\sum_{\ell,\eta}\langle\varepsilon_{\alpha\ell}^{-\eta}(k)|{\cal S}|\varepsilon_{\alpha\ell}^{\eta}(k)\rangle\langle\varepsilon_{\alpha\ell}^{\eta}(k)|i\partial_{k}|\varepsilon_{\alpha\ell}^{-\eta}(k)\rangle,\label{eq:WAapp2}
\end{alignat}
where we have used the fact that $\langle\varepsilon_{\alpha\ell}^{\eta}(k)|{\cal S}|\varepsilon_{\alpha\ell'}^{\eta'}(k)\rangle\propto\delta_{\ell\ell'}\delta_{\eta,-\eta'}$.
Similarly, we can express the MCD in Eq.~(\ref{eq:C12}) in the time frame $\alpha$ as:
\begin{alignat}{1}
c_{\alpha}= & \lim_{M\rightarrow\infty}\frac{1}{M}\sum_{m=1}^M\int_{-\pi}^{\pi}\frac{idk}{2\pi}\sum_{\ell,\ell'=1,2}\sum_{\eta,\eta'=\pm}\nonumber \\
\times & {\rm Tr}\{{\cal S}e^{-im\varepsilon_{\ell}^{\eta}(k)}|\varepsilon_{\alpha\ell}^{\eta}(k)\rangle\langle\varepsilon_{\alpha\ell}^{\eta}(k)|\nonumber \\
\times & \partial_{k}[e^{-im\varepsilon_{\ell'}^{\eta'}(k)}|\varepsilon_{\alpha\ell'}^{\eta'}(k)\rangle\langle\varepsilon_{\alpha\ell'}^{\eta'}(k)|]\},\label{eq:CAapp1}
\end{alignat}
where $\ell=1,2$ and $\eta=\pm$. Taking the trace explicitly in
the quasienergy eigenbasis, we can further express $c_{\alpha}$
as
\begin{alignat}{1}
c_{\alpha}= & \lim_{M\rightarrow\infty}\frac{1}{M}\sum_{m}\int_{-\pi}^{\pi}\frac{dk}{2\pi}\sum_{\ell,\eta}\label{eq:CAapp2}\\
\times & \left[e^{-i2m\varepsilon_{\ell}^{\eta}(k)}m\partial_{k}\varepsilon_{\ell}^{\eta}(k)\langle\varepsilon_{\alpha\ell}^{\eta}(k)|{\cal S}|\varepsilon_{\alpha\ell}^{\eta}(k)\rangle\right.\nonumber \\
+ & \langle\varepsilon_{\alpha\ell}^{-\eta}(k)|{\cal S}|\varepsilon_{\alpha\ell}^{\eta}(k)\rangle\langle\varepsilon_{\alpha\ell}^{\eta}(k)|i\partial_{k}|\varepsilon_{\alpha\ell}^{-\eta}(k)\rangle\nonumber \\
+ & \left.e^{-i2m\varepsilon_{\ell}^{\eta}(k)}\langle\varepsilon_{\alpha\ell}^{-\eta}(k)|{\cal S}|\varepsilon_{\alpha\ell}^{\eta}(k)\rangle\langle i\partial_{k}\varepsilon_{\alpha\ell}^{\eta}(k)|\varepsilon_{\alpha\ell}^{-\eta}(k)\rangle\right].\nonumber 
\end{alignat}
It is clear that the term in the second line of Eq.~(\ref{eq:CAapp2})
vanishes, since $\langle\varepsilon_{\alpha\ell}^{\eta}(k)|{\cal S}|\varepsilon_{\alpha\ell}^{\eta}(k)\rangle\propto\langle\varepsilon_{\alpha\ell}^{\eta}(k)|\varepsilon_{\alpha\ell}^{-\eta}(k)\rangle=0$.
The term in the fourth line of Eq.~(\ref{eq:CAapp2}) carries an oscillating
phase factor $e^{-i2m\varepsilon_{\ell}^{\eta}(k)}$. So it will also
vanish under the long-time average $\lim_{M\rightarrow\infty}\frac{1}{M}\sum_{m}$.
The only contribution to the value of $c_{\alpha}$ left after taking the long-time
average comes from the term in the third line of Eq.~(\ref{eq:CAapp2}),
which is independent of the driving period index $m$. Working out
the sum over $m$, we end with
\begin{equation}
c_{\alpha}=\int_{-\pi}^{\pi}\frac{dk}{2\pi}\sum_{\ell,\eta}\langle\varepsilon_{\alpha\ell}^{-\eta}(k)|{\cal S}|\varepsilon_{\alpha\ell}^{\eta}(k)\rangle\langle\varepsilon_{\alpha\ell}^{\eta}(k)|i\partial_{k}|\varepsilon_{\alpha\ell}^{-\eta}(k)\rangle.\label{eq:CAapp3}
\end{equation}
Therefore, comparing Eqs.~(\ref{eq:WAapp2}) with (\ref{eq:CAapp3}),
we conclude that the winding number and MCD in the symmetric time
frame $\alpha$ satisfy $w_{\alpha}=c_{\alpha}$ for $\alpha=1,2$.
Finally, according to the definitions of topological invariants $(w_{0},w_{\pi})\equiv(\frac{w_{1}+w_{2}}{2},\frac{w_{1}-w_{2}}{2})$
and dynamical winding numbers $(c_{0},c_{\pi})\equiv(\frac{c_{1}+c_{2}}{2},\frac{c_{1}-c_{2}}{2})$
in the main text, we arrive at the desired relationship between the topological
and dynamical winding numbers of the PQSCL, i.e., $(w_{0},w_{\pi})=(c_{0},c_{\pi})$
as given by Eq.~(\ref{eq:WC0P}) in the main text.

\end{document}